\documentclass[10pt]{iopart}
\usepackage[breaklinks=true,colorlinks,citecolor=blue,linkcolor=blue,urlcolor=blue]{hyperref}
\usepackage{bm}
\usepackage{iopams} 
\expandafter\let\csname equation*\endcsname\relax

\expandafter\let\csname endequation*\endcsname\relax

\usepackage{amsmath}
\usepackage{graphicx,graphics}
\usepackage[frozencache,cachedir=.]{minted} 
\begin{document}
\title{Diffusive solver: a diffusion-equations solver based on FEniCS}
\author{Iacopo Torre}
\ead{iacopo.torre@icfo.eu}
\address{ICFO-Institut de Ci\`{e}ncies Fot\`{o}niques, The Barcelona Institute of Science and Technology, Av. Carl Friedrich Gauss 3, 08860 Castelldefels (Barcelona),~Spain}
\vspace{10pt}
\begin{abstract}
Many steady-state transport problems in condensed matter physics can be reduced to a set of coupled diffusion equations.
This is true in particular when relaxation processes are sufficiently fast that the system is in the diffusive --opposite of ballistic-- regime.
Here we describe a python package \cite{repo}, based on FEniCS \cite{FEniCS}, that solves this type of problems with an arbitrary number degrees of freedom that can represent charge, spin, energy, band or valley flavours.
Generalized conductivities and responsivities, characterizing completely the linear response of the system to external biases and sources, are automatically computed from the solutions.
We solve two simple example of magneto-transport and thermoelectric transport for illustrative purpose. 
\end{abstract}
%
%
\section{Introduction}
Transport in many-particle systems can be typically described, at steady-state, and if the size of the system is large with respect to the mean-free-path of the individual particles, by a set of diffusion equations for few collective variables.
The simplest example is given by the local form of the Ohm-s law 
\begin{equation}\label{Ohm}
\bm J(\bm r) = -\bm \sigma(\bm r) \nabla V(\bm r),
\end{equation}
where $\bm J(\bm r)$ is the electric current density, $\bm \sigma(\bm r)$ the conductivity tensor, and $V(\bm r)$ the electric potential.
This, combined with the steady-state continuity equation, expressing charge conservation
\begin{equation}
\nabla \cdot \bm J(\bm r) =0,
\end{equation}
yields the equation for the potential
\begin{equation}\label{diffusion_potential}
-\nabla \cdot[\bm \sigma(\bm r) \nabla V(\bm r)] = 0.
\end{equation}
Equation \ref{diffusion_potential} (where $\bm \sigma(\bm r) >0$ in the matrix sense) is the simplest example of diffusion-like equation.
Since charge is not created and does not decay there are no sources and the associated decay length is infinite in this simple example.
In a typical device configuration we want to solve (\ref{diffusion_potential}) with boundary conditions specifying the potential fixed by external sources on some parts of the device boundary, i.e. the contacts, and conservation of charge ($\hat{n} \cdot \bm J = 0$) on the rest of the boundary.

The Finite Element Method (FEM) is a numerical technique to solve partial differential equations (See for example \cite{FEM} for an introduction to FEM).
It is based on finding a {\it weak} solution of the differential equation in a finite dimensional subspace of the solution space given by piecewise polynomial functions.
FEniCS \cite{FEniCS,FEniCS_book} is an open-source (LGPLv3 license) computing platform for solving partial differential equations (PDEs) using FEM, provided with an high-level Python interface.

We developed the python package \mintinline{python}{diffusive_solver} \cite{repo} (released under the same LGPLv3 license), based on FEniCS, that allows a quick solution of coupled diffusion equations (See Sect. \ref{math} for the precise mathematical definition of the problem) in arbitrary device geometries in one (1D), two (2D) and three (3D) spatial dimensions (although internal meshing and plotting functionalities are available only in 2D). 
While its scope is much narrower than that of the full FEniCS library its focus is on the simplicity of use for a user with no experience with the underlying FEniCS library.

These notes are organized as follows.
In Sect. \ref{math} we give a precise mathematical description of the general problem that can be solved with \mintinline{python}{diffusive_solver}.
An elementary mageto-transport example is presented in Sect. \ref{example_conduction}, while a more advanced example of thermoelectric transport is presented in Sect. \ref{example_theroelectric} to clarify the main features of the package in a concrete framework. The results and perspectives for future development are summarized in Sect. \ref{summary}. Appendices are devoted to the demonstration of the key mathematical results.
\section{Mathematical formulation}
\label{math}
The most general {\it linear} diffusion equation problem can be formulated, in a connected domain $\Omega\in \mathbb{R}^{N_d}$, in terms of $N_f$ fields $\phi_{\alpha}(\bm r)$ with $\alpha =0,\dots,N_f-1$ and $N_f$ corresponding currents $\bm J_{\alpha}(\bm r)$, respecting generalized continuity equations
\begin{equation}\label{continuity}
\sum_i \partial_iJ_{\alpha, i}(\bm r)=
-\sum_\beta\Gamma_{\alpha\beta}(\bm r)\phi_\beta(\bm r)+F_{\alpha}(\bm r).
\end{equation}
Here, $\partial_i$ denotes the partial derivative with respect to the i-th spatial direction, greek indices (ranging from $0$ to $N_f-1$)identify different field components, and latin 
indices (ranging from $0$ to $N_d-1$, $N_d$ being the number of spatial dimensions) spatial components, $\Gamma_{\alpha\beta}(\bm r)$ is a position-dependent, {\it semi-positive-defined}, relaxation matrix in the field indices $\alpha, \beta$ and $F_{\alpha}(\bm r)$ is a vector of source terms. 

The currents are in turn related to the fields by a linear constitutive relation
\begin{equation}\label{constitutive}
J_{\alpha, i}(\bm r)\equiv -\sum_{\beta j}
L_{\alpha\beta,ij}(\bm r)\partial_j\phi_\beta(\bm r),
\end{equation}
where $L_{\alpha\beta,ij}(\bm r)$ is a matrix in both field and Cartesian indices that must be {\it positive defined}.
Substitution of (\ref{constitutive}) into (\ref{continuity}) leads to the system of PDEs
\begin{equation}\label{system}
-\sum_{\beta ij}\partial_i[L_{\alpha\beta,ij}(\bm r)\partial_j\phi_\beta(\bm r)]
+\sum_{\beta}\Gamma_{\alpha\beta}(\bm r)\phi_\beta(\bm r)
=F_{\alpha}(\bm r).
\end{equation}

The system of equations (\ref{system}) has a unique solution once supplemented with suitable boundary conditions on $\partial \Omega$, the boundary of $\Omega$. 
The boundary contains $N_c$ regions $C_1,\dots ,C_{N_c}$ dubbed contacts. 
In each of these regions we impose Robin boundary conditions in the form
\begin{equation}
\phi_\alpha(\bm r)- \bar{R}_{\alpha, n}\sum_i J_{\alpha, i}(\bm r) \hat{n}_i(\bm r)  = V_{\alpha, n} ,\quad \mbox{if } \bm r \in C_n,
\end{equation} 
where $V_{\alpha, n}$ are generalized biases, and $\bar{R}_{\alpha, n}$ are generalized, {\it semi-positive}, contact resistances, and $\hat{\bm n}(\bm r)$ is the outward normal unit vector.
Note that if the contact resistances are zero the above condition reduces to the more standard Dirichlet boundary condition
\begin{equation}
\phi_\alpha(\bm r) = V_{\alpha, n}\quad \mbox{if }, \bm r \in C_n.
\end{equation}
In the remaining part of the boundary we impose homogeneous Neumann boundary conditions 
\begin{equation}
\sum_i J_{\alpha, i}(\bm r) \hat{n}_i(\bm r) = 0 ,\quad \mbox{if } \bm r \in \partial\Omega \setminus \cup_n C_n,
\end{equation}
corresponding to vanishing current leaving the domain.

Under these boundary conditions the solution of (\ref{system}) is unique as demonstrated in \ref{proof_uniqueness}.

\subsection{Basis functions}
Since the PDE problem is linear its most general solution can be written as a linear combination of basis functions that are solutions of the homogeneous problem ($F\equiv 0$) plus a solution of the inhomogeneous equation.
We define a basis of $N_c \cdot N_f$ functions $\Phi^{(\alpha,n)}_\beta(\bm r)$ satisfying the homogeneous problem
\begin{equation}\label{system_homo}
-\sum_{\beta,ij}\partial_i[L_{\gamma\beta,ij}(\bm r)\partial_j\Phi^{(\alpha,n)}_\beta(\bm r)]
+\sum_{\beta}\Gamma_{\gamma\beta}(\bm r)\Phi^{(\alpha,n)}_\beta(\bm r), 
=0,
\end{equation}
with the biases $V_{\beta, m} = \delta_{mn} \delta_{\alpha \beta}$, i.e. all the biases set to zero except the $\alpha$ component of the bias on the contact $n$ that is set to $1$.

We also define the inhomogeneous solution $\Phi^{F}_\beta(\bm r)$ as the solution
\begin{equation}\label{system_inhomo}
-\sum_{\beta,ij}\partial_i[L_{\gamma\beta\,ij}(\bm r)\partial_j\Phi^{F}_\beta(\bm r)]
+\sum_{\beta}\Gamma_{\gamma\beta}(\bm r)\Phi^{F}_\beta(\bm r), 
=F_\gamma(\bm r),
\end{equation}
with {\it all} the biases set to zero.
At each of these solutions is associated a current field
\begin{equation}
J_{\gamma,i}^{(\alpha,n)}(\bm r) = -\sum_{\beta,j} L_{\gamma\beta,ij}(\bm r) \partial_j\Phi^{(\alpha,n)}_\beta(\bm r),
\end{equation}
and 
\begin{equation}
J_{\gamma,i}^{F}(\bm r) = -\sum_{\beta,j} L_{\gamma\beta,ij}(\bm r) \partial_j\Phi^{F}_\beta(\bm r),
\end{equation}

In this way the solution $\phi_\beta (\bm r)$ with generic biases $V_{\alpha,n}$ and source term $F$ can be expressed as
\begin{equation}\label{decomposition}
\phi_\beta (\bm r) = \sum_{\alpha, n} V_{\alpha,n} \Phi^{(\alpha,n)}_\beta(\bm r) + \Phi^{F}_\beta(\bm r).
\end{equation}

\subsection{Adjoint problem}
The adjoint problem of (\ref{system}) is defined by
\begin{equation}\label{system_adj}
-\sum_{\beta,ij}\partial_i[L_{\alpha\beta,ij}^{T}(\bm r)\partial_j\phi_\beta(\bm r)]
+\sum_{\beta}\Gamma_{\alpha\beta}^{T}(\bm r)\phi_\beta(\bm r)
=F_{\alpha}(\bm r),
\end{equation}
where the transposed matrices $L_{\alpha\beta,ij}^{T}(\bm r) = L_{\beta\alpha,ji}(\bm r)$ and $\Gamma_{\alpha\beta}^{T}(\bm r)= \Gamma_{\beta\alpha}(\bm r)$.
Note that the transpose acts on {\it both} field {\it and} Cartesian indices.
As done for the direct problem we define basis functions $\tilde{\Phi}^{(\alpha,n)}_\beta(\bm r)$ as the solutions of (\ref{system_adj})
with $F\equiv 0$ and $V_{\beta, m} = \delta_{mn} \delta_{\alpha \beta}$.
The adjoint inhomogeneous solution is instead defined as $\tilde{\Phi}^{F}_\beta(\bm r)$, the solution of (\ref{system_adj}) with all the biases $V_{\beta, m}$ set to zero.
Again, we associate to these solutions the corresponding current fields
\begin{equation}
\tilde{J}_{\gamma,i}^{(\alpha,n)}(\bm r) = -\sum_{\beta,j} L_{\gamma\beta,ij}^T(\bm r) \partial_j\Phi^{(\alpha,n)}_\beta(\bm r),
\end{equation}
and 
\begin{equation}
\tilde{J}_{\gamma,i}^{F}(\bm r) = -\sum_{\beta,j} L_{\gamma\beta,ij}^T(\bm r) \partial_j\Phi^{F}_\beta(\bm r).
\end{equation}
A problem that is equal to its adjoint because the matrices $L$ and $\Gamma$ are symmetric is self-adjoint. 

\subsection{Fluxes and linear response}
Beside the spatial dependence of the fields and currents, an important quantity we want to solve for is the flux of the current ${\bm J}_{\alpha}(\bm r)$ at the contact $m$, defined by 
\begin{equation}
I_{\alpha,m}\equiv \int_{C_m} \sum_i\hat{n}_i(\bm r) J_{\alpha, i}(\bm r)ds.
\end{equation}
Note that we chose the positive direction of each flux to be the one exiting the device.

The basis function decomposition (\ref{decomposition}) allows us to express the fluxes as a linear combination of biases and sources according to
\begin{equation}
I_{\alpha,m} = \sum_{\beta, n} G_{\alpha,m;\beta,n}V_{\beta, n} + S_{\alpha,m}[F],
\end{equation}
where $G_{\alpha,m;\beta,n}$ does not depend on $F$ and $S_{\alpha\,m}$ is a linear function of $F$.
More explicitly, the linear response matrix $G_{\alpha,m;\beta,n}$ is defined by
\begin{equation}
G_{\alpha,m;\beta,n}  \equiv \int_{C_m} \sum_i\hat{n}_i(\bm r) J_{\alpha, i}^{(\beta,n)}(\bm r)ds,
\end{equation}
while
\begin{equation}\label{S_def}
S_{\alpha,m}[F]  \equiv \int_{C_m} \sum_i\hat{n}_i(\bm r) J_{\alpha, i}^{F}(\bm r)ds.
\end{equation}
The response matrix has the following two properties whose demonstration is provided in \ref{symmetry}.
First,
\begin{equation}
G_{\alpha,m;\beta,n}=\tilde{G}_{\beta,n;\alpha,m},
\end{equation}
where 
\begin{equation}
\tilde{G}_{\alpha,m;\beta,n}  \equiv \int_{C_m} \sum_i\hat{n}_i(\bm r) \tilde{J}_{\alpha, i}^{(\beta,n)}(\bm r)ds,
\end{equation}
is the response matrix of the adjoint problem.

Second, 
\begin{equation}
\sum_{\alpha \beta ,nm}V_{\alpha,m}G_{\alpha,m;\beta,n}V_{\beta,n}\leq 0
\end{equation}
for every choice of $V_{\alpha,m}$, the equality holding {\it only} if all the $V_{\alpha,m}$ vanish.
This means that $G$ is negative defined.

The source term $S$ can be rewritten in terms of generalized {\it responsivities} as
\begin{equation}\label{S_resp}
S_{\alpha,m}[F] = \sum_{\beta}\int_{\Omega}  {\cal R}^{(\alpha, m)}_\beta(\bm r) F_{\beta}(\bm r)d\bm r.
\end{equation}
As a consequence of the reciprocity principle \cite{MarkReciprocity, Song} the responsivities are simply the solutions of the adjoint problem 
\begin{equation}\label{reciprocity}
{\cal R}^{(\alpha, m)}_\beta(\bm r) = \tilde{\Phi}_\beta^{(\alpha, m)}(\bm r),
\end{equation}
as demonstrated in \ref{app_reciprocity}.

The most general way of expressing the response to biases and sources is therefore
\begin{equation}
I_{\alpha,m} =\sum_{\beta, n} G_{\alpha,m;\beta,n}V_{\beta, n} + \sum_{\beta}\int_{\Omega} {\cal R}^{(\alpha, m)}_\beta(\bm r)F_{\beta}(\bm r)d\bm r .
\end{equation}
Our package allows to solve for the solution for an arbitrary configuration of biases and source terms as well as the automatic computation of the  matrices $G_{\alpha,m;\beta,n}$ and $S_{\alpha,m}$ and the generalized responsivities ${\cal R}^{(\alpha, m)}_\beta(\bm r)$.
\subsection{Discretization of the PDE problem}
The weak formulation of the problem is obtained by multiplying (\ref{system}) by a test function $u_{\alpha}(\bm r)$ and integrating over $\Omega$. Following the usual procedure of integrating by parts the term with the derivatives, and making use of the homogeneous Neumann conditions outside the contacts we obtain
\begin{equation}
\begin{split}
& \int_{\Omega} \left[\sum_{\alpha\beta ij}\partial_i u_\alpha(\bm r)L_{\alpha\beta,ij}(\bm r)\partial_j\phi_\beta(\bm r)
+\sum_{\alpha\beta} u_\alpha(\bm r)\Gamma_{\alpha\beta}(\bm r)\phi_\beta(\bm r)- \sum_\alpha  u_\alpha(\bm r)F_{\alpha}(\bm r)\right] d{\bm r}=\\
& = \sum_{m, \alpha}\sum_{\beta, ij} \int_{C_m} \hat{n}_i (\bm r) u_\alpha(\bm r) L_{\alpha\beta,ij}(\bm r)\partial_j\phi_\beta(\bm r) ds.
\end{split}
\end{equation}
The terms of the sum on the right hand side evaluate to
\begin{equation}
\int_{C_m}  u_\alpha (\bm r)\frac{V_{\alpha,m} - \phi_\alpha(\bm r)}{\bar{R}_{\alpha, m}}ds,
\end{equation}
for the indices $\alpha, m$ corresponding to non-zero contact resistances.
For the values $\alpha, m$ corresponding to a vanishing $\bar{R}_{\alpha, m}$ an intrinsic Dirichlet boundary condition is imposed.
This implies that the $\alpha$ component of the test function $u_\alpha(\bm r)$ vanishes at the contact $m$ causing the corresponding integral to vanish.
 
The discrete problem is obtained by expanding the fields on continuous, piecewise-linear (Lagrange) finite elements.
All the coefficients are considered as constants within each cell by evaluating the corresponding functions at the cell centres, i.e. they are integrated using a 0-th order quadrature inside each cell.
Coherently, during post-processing, the currents are expanded on discontinuous piecewise-constant elements.
\section{Example: Electrical conduction in magnetic field}
\label{example_conduction}
The first example we provide is the solution of Eq. (\ref{diffusion_potential}) with the conductivity tensor given by
\begin{equation}
\bm \sigma (\bm r) = \begin{pmatrix}\sigma & \sigma_{\rm H}\\-\sigma_{\rm H} & \sigma \end{pmatrix},
\end{equation}
$\sigma$, being the normal conductivity and $\sigma_{\rm H}$ being the Hall conductivity, assumed to be spatially uniform.

After installing FEniCS \cite{FEniCS}, our package can be installed from the python package index as
\begin{minted}[mathescape]{shell}
pip install  diffusive-solver
\end{minted}
or downloaded from the project repository \cite{repo}.

We start by importing our package using
\begin{minted}[mathescape]{python}
from diffusive_solver import *
\end{minted}

\subsection{Geometry definition}
Next, we define a sample geometry.
For illustrative purposes we use just a rectangular geometry with contacts placed on two sides.
A quick way to define a simple polygonal geometry is the following
\begin{minted}[mathescape]{python}
geometry = Geometry.from_points(points_coords =[[-3,1],[-3,0],[3,0],[3,1]], 
                                contacts_positions = [[0,1],[2,3]], 
                                resolution = 20)
\end{minted}
where \mintinline{python}{points_coords} are the vertices of the polygon ordered in a counterclockwise direction, \mintinline{python}{contacts_positions} is a list of length $N_{\rm c}$ whose element are lists of the vertices enclosed in each contact (that must be contiguous).
In this example the first contact stretches from the point \mintinline{python}{0} (\mintinline{python}{[-3,1]}) to the point \mintinline{python}{1} (\mintinline{python}{[-3,0]}).

If the geometry is more complicated (but still polygonal) it can be convenient to put the coordinates in a file \mintinline{python}{'rectangle.txt'} with the following sintax
\begin{minted}[mathescape]{python}
#X[um]	Y[um]	contact
-3.		1.		1
-3.		0.		1
 3.		0.		2
 3.		1.		2
\end{minted}
and then build the geometry using the command
\begin{minted}[mathescape]{python}
geometry = Geometry.from_txt_file(filename = 'rectangle.txt', 
                                  resolution = 20)
\end{minted}
In both cases increasing the parameter \mintinline{python}{resolution} will give a finer mesh.

A third option, that is strongly recommended for complex geometries, and mandatory for 1D and 3D problems, is to use an external mesh generator like GMSH \cite{GMSH} to produce a .msh file that can then be read using
\begin{minted}[mathescape]{python}
geometry = Geometry.from_msh_file(filename = 'rectangle.msh',
                                  dim = 2, dim_keep = 2, n_contacts = 2)
\end{minted}
Here \mintinline{python}{dim} is the topological dimension of the mesh while \mintinline{python}{dim_keep} (default value \mintinline{python}{dim}) specifies the dimensionality of the space the mesh could be embedded in. One dimensional problems need to be embedded in two dimensions, with the last coordinate equal to zero.

The contacts must be marked as physical entities in the mesh and numbered from $1$ to $n_{\rm c}$.
If this option is used the presence of physical subdomains in the mesh is detected and stored in a cell function available as \mintinline{python}{geometry.subdomain_marker}.

Geometrical properties of the mesh are stored as Numpy arrays to make them easily accessible for external post-processing.
More precisely, \mintinline{python}{geometry.coordinates} is a $N_{\rm vert} \times N_{\rm d}$ array containing the coordinates of the $N_{\rm vert}$ vertices of the mesh, \mintinline{python}{geometry.cells_centers} is a $N_{\rm cells} \times N_{\rm d}$ array containing the coordinates of the centres of the $N_{\rm cells}$ cells of the mesh, and \mintinline{python}{geometry.cells} is a $N_{\rm cells} \times N_{\rm d}+1$ array of integers containing the cells connectivity.

The class \mintinline{python}{Geometry} also provides methods for rapidly checking the correctness of the geometry.
The command \mintinline{python}{geometry.plot()} plots the mesh and the marked contacts, \mintinline{python}{geometry.plot_subdomains()} plots the subdomains (if present) and \mintinline{python}{geometry.check_dimensions()} prints the areas and volumes of the different parts of the geometry.

\subsection{Coefficients definition}
The next step is to define the coefficients of the problem.
The class \mintinline{python}{Matrix_Expression} provides a simple interface to define them.
In our example the coefficients are just constant. We will see a more advanced usage in the next example.
First we create the matrix $L$ in this way
\begin{minted}[mathescape]{python}
sigma = 1.
sigma_H = 1.
L_00 = Matrix_Expression([[sigma, sigma_H],
                          [-sigma_H,sigma]], 
                         scalar = False, dimension = 2)
L = [[L_00]]
\end{minted}
note that \mintinline{python}{L_00} is a $2 \times 2$ matrix in the spatial indices, while \mintinline{python}{L} is a list of lists, with the two dimensions of length one because we only have one field.
The $\Gamma$ and $F$ matrices are null matrices of appropriate dimensions. Here we define them explicitly using
\begin{minted}[mathescape]{python}
null_mat = Matrix_Expression(0, scalar = True, dimension = 1)
Gamma = [[null_mat]]
F = [null_mat]
\end{minted}
where the first command creates a null matrix of spatial dimension one.
Note that if  \mintinline{python}{dimension} and \mintinline{python}{scalar} are not specified the default is a scalar matrix of dimension one, that is a real number.
\subsection{Problem solution}
Now we can define the PDE problem using the geometry and the coefficients
\begin{minted}[mathescape]{python}
problem = Problem(geometry = geometry,  
                  n_fields = 1,
                  self_adjoint = False,
                  L = L,
                  Gamma = Gamma,
                  contact_resistances = [[0.1, 0.1]],
                  biases = [[1.,0.]],
                  F = F)
\end{minted}
Here we passed explicitly all the arguments for the sake of clarity. However, omitting \mintinline{python}{n_fields}, \mintinline{python}{self_adjoint}, \mintinline{python}{Gamma}, and \mintinline{python}{F}, would have led to the same result because of default options. 
The solution of the problem can be triggered using
\begin{minted}[mathescape]{python}
problem.solve()
\end{minted}
while the current fields corresponding to the solution are calculated using
\begin{minted}[mathescape]{python}
problem.compute_currents()
\end{minted}
The fields and currents are available as lists of FEniCS functions as \mintinline{python}{problem.fields} and \mintinline{python}{problem.currents} respectively.

The results of the computation, (fields and currents if computed) can be saved in .xdmf format using
\begin{minted}[mathescape]{python}
problem.save(folder = 'output_folder')
\end{minted}
The results can be easily visualized with ParaView \cite{ParaView}.
The fluxes and the response matrices $G$ and $S$ can be printed using
\begin{minted}[mathescape]{python}
print(problem.fluxes)
print(problem.response_matrix)
print(problem.source_vector)
\end{minted}
All the parameters in the problem definition are fixed (another instance must be created to solve a problem with different coefficients) with the exception of the source terms \mintinline{python}{biases} and \mintinline{python}{F}.
These can be changed by solving again the problem with
\begin{minted}[mathescape]{python}
problem.solve(biases = new_biases, F = new_F)
\end{minted}
\subsection{Post-processing}
\begin{figure}
\includegraphics[scale=0.75]{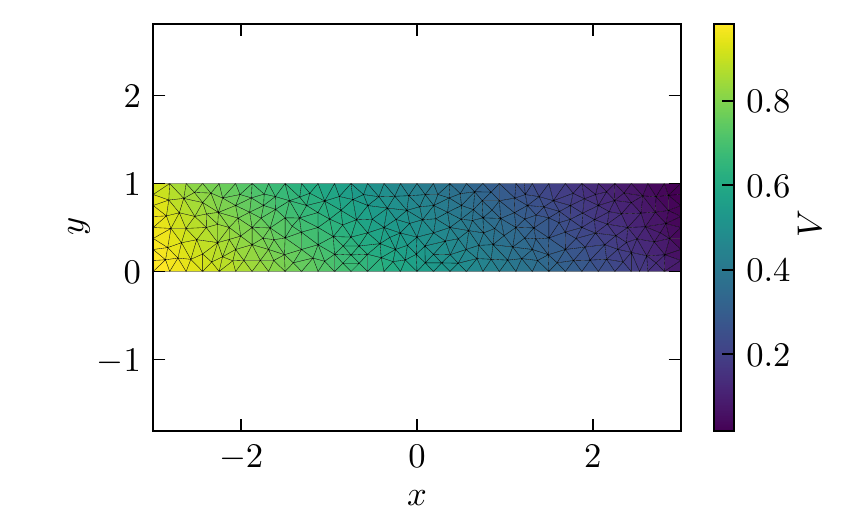}
\includegraphics[scale=0.75]{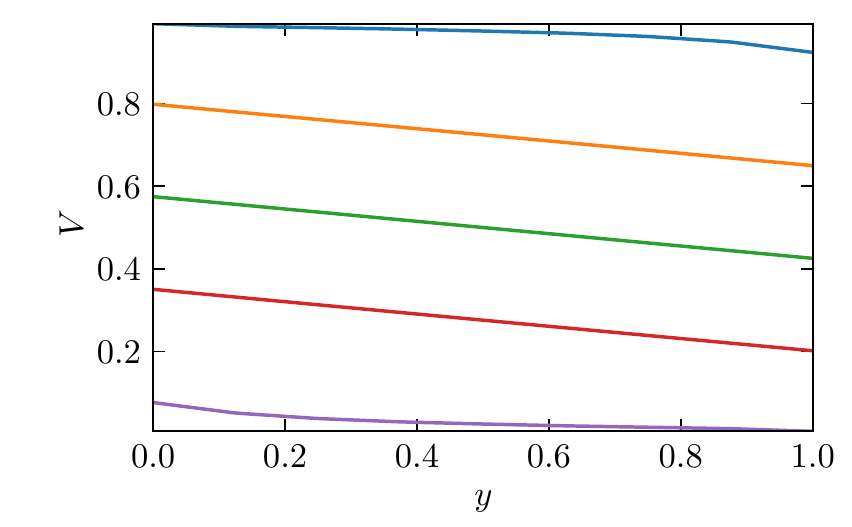}
\caption{\label{fig1} Left. Potential map of the solution of the magnetotransport problem. Right. Cross sections of the solution taken at $x=-3$ (blue), $x=-1.5$ (orange), $x=0$ (green), $x=1.5$ (red), $x=3$ (violet).}
\end{figure}

We can directly plot a field component making use of 
\begin{minted}[mathescape]{python}
problem.plot_field(field = 0)
plt.axis('equal')
cb = plt.colorbar()
cb.set_label('$V$')
plt.xlabel('$x$')
plt.ylabel('$y$')
\end{minted}
The result is shown in Fig. \ref{fig1}
We can also evaluate the field at specific positions and plot line cuts (see the results in Fig. \ref{fig1})
\begin{minted}[mathescape]{python}
y= np.linspace(0,1)
for xi in np.linspace(-3, 3, num = 5):
    plt.plot(y,[problem.fields[0](xi,yi) for yi in y])
plt.xlabel('$y$')
plt.ylabel('$V$')
\end{minted}
Evaluating the fields (or the currents) is particularly useful to obtain the solution on a regular structured mesh for further processing.
Finally, the values of fields at vertices and currents at cell centres are available as Numpy arrays for post-processing.
The property \mintinline{python}{problem.fields_vertices} is a $N_{\rm f}\times N_{\rm vert}$ array containing the values of the fields at mesh vertices, \mintinline{python}{problem.currents_cells} is a $N_{\rm f}\times N_{\rm cells}\times N_{\rm d}$ containing the values of currents at cells centres.
The ordering of vertices and cells is coherent with the ones used in the geometrical properties.
\section{Example: thermoelectric transport}
\label{example_theroelectric}
Here we discuss a more elaborate example that involves the solution of two coupled fields.
In inhomogeneous conductors the charge and heat transport are coupled by the Seebeck-Peltier effect \cite{AshcroftMermin}.
This means that a gradient of the temperature $T(\bm r)$ can induce an electric current and a gradient of potential $V(\bm r)$ can induce a heat current, according to
\begin{equation}
\begin{pmatrix}
\bm J(\bm r)\\
\bm q(\bm r)
\end{pmatrix}
=-
\begin{pmatrix}
\sigma(\bm r) &  T_0\sigma (\bm r) S(\bm r)\\
T_0\sigma (\bm r)S(\bm r) & T_0^2\sigma(\bm r) S^2(\bm r) +T_0\kappa(\bm r)
\end{pmatrix}
\begin{pmatrix}
\nabla V(\bm r)\\
T_0^{-1}\nabla \delta T(\bm r)
\end{pmatrix},
\end{equation}
where $\bm J(\bm r)$ is the electric current, $\bm q(\bm r)$ is the heat current, $\sigma(\bm r)$ is the electrical conductivity, $\kappa(\bm r)$ the thermal conductivity, $S(\bm r)$ the Seebeck coefficient, $\delta T(\bm r) \equiv T(\bm r)-T_0$, is the local temperature variation $T_0$ being the equilibrium temperature. Note that we put the equations in a form in which the matrix $L$ is explicitly symmetric thanks to Onsager relations.
In a two-dimensional sample charge is conserved but heat can be lost to the substrate to a coupling $g(\bm r)$ \cite{ThermoelectricDetection}. We therefore have the continuity equations.
\begin{equation}
\begin{pmatrix}
\nabla \cdot \bm J(\bm r)\\
\nabla \cdot \bm q(\bm r)
\end{pmatrix}
+
\begin{pmatrix}
0 &  0\\
0 & T_0g(\bm r)
\end{pmatrix}
\begin{pmatrix}
 V(\bm r)\\
T_0^{-1}\delta T(\bm r)
\end{pmatrix}
=
\begin{pmatrix}
0\\
Q(\bm r)
\end{pmatrix}.
\end{equation}
\subsection{Coefficients definition}
The definition of the geometry proceeds exactly as in the previous example but, in this case, we define spatially varying coefficients.
We define the charge-temperature coupling (the upper right corner of the matrix $L$) as a smooth step function representing, for example, a pn junction. An object \mintinline{python}{Matrix_Expression} can be initialized with a python function (or callable) or a matrix of functions specifying the matrix elements as 
\begin{minted}[mathescape]{python}
ct = Matrix_Expression([[lambda x, y: np.tanh(x),lambda x, y: 0.],
                        [lambda x, y: 0.,lambda x, y: np.tanh(x)]], 
                       dimension = 2, scalar = False, mesh = geometry.mesh)
\end{minted}
Note that in this case the mesh over which the functions must be calculated needs to be passed explicitly.
We define the other coefficients as constant
\begin{minted}[mathescape]{python}                            
cc = Matrix_Expression([[1.,0],
                        [0,1.]], 
                       dimension = 2, scalar = False)
tt = Matrix_Expression(2.)
L = [[cc,ct],
     [ct,tt]]
\end{minted}
and similarly, for the matrices $\Gamma$ and $F$
\begin{minted}[mathescape]{python} 
Gamma = [[Matrix_Expression(0),Matrix_Expression(0)],
         [Matrix_Expression(0),Matrix_Expression(5)]]

F = [Matrix_Expression(0), 
     Matrix_Expression(lambda x, y: np.exp(-10*(x-1)**2-10*(y-0.2)**2), 
                       mesh = geometry.mesh, dimension = 1)]
\end{minted}
Here the source $Q(\bm r)$ represents the heat introduced in the system by an illuminating Gaussian light beam.
Another way of defining coefficients that is very useful in the case of piecewise constant coefficients representing, for example, different materials is using a marker.
If the geometry was created from an external mesh with marked subdomains we can define a piecewise constant matrix with
\begin{minted}[mathescape]{python}
ct = Matrix_Expression({1: [[-1.,0],[0,-1]], 
                        2: [[1.,0],[0,1]]}, 
                        marker = geometry.subdomain_marker,
                        dimension = 2, scalar = False)
\end{minted}
where the first line applies to the subdomain 1 and the second to subdomain 2.
For further examples on more complex geometries we refer to the example notebooks in the project repository \cite{repo}.
\subsection{Problem solution}
\begin{figure}
\includegraphics[scale=0.75]{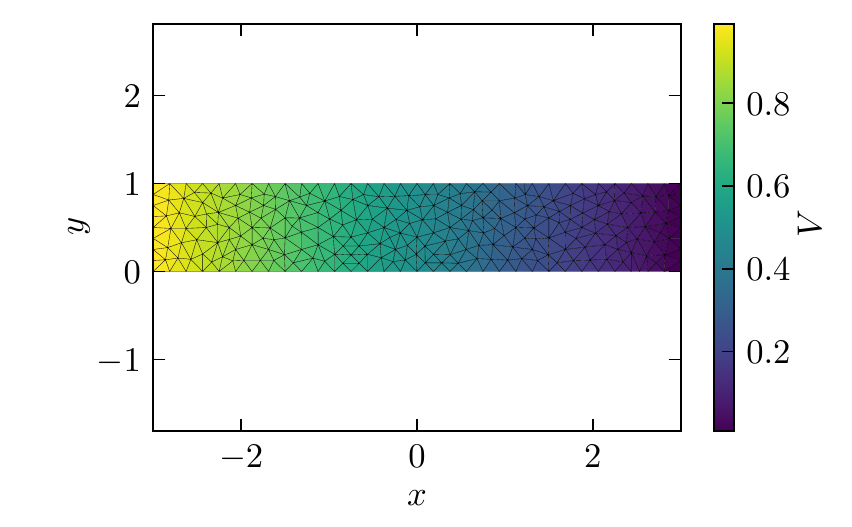}
\includegraphics[scale=0.75]{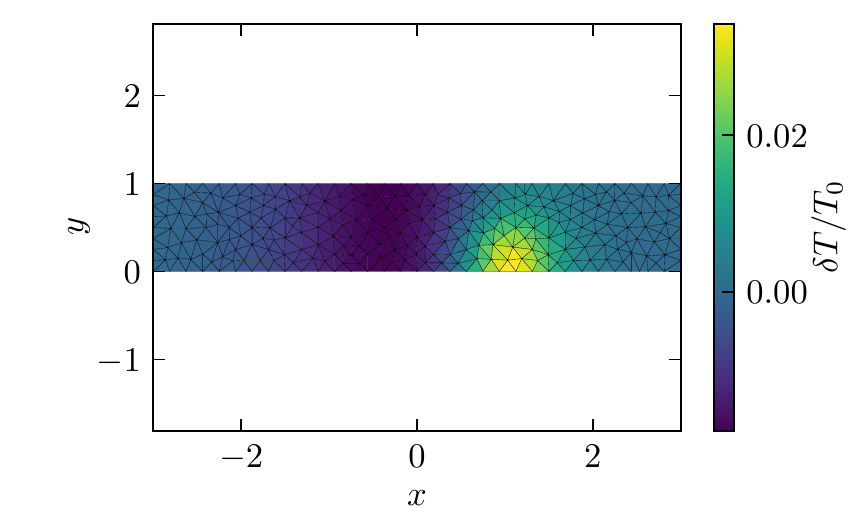}
\caption{\label{fig2} Left. Voltage field obtained for the thermoelectric example. Right. Temperature field obtained for the same example. Note the impact of the Seebeck gradient extending away from the junction and the temperature rise due to the light spot.}
\end{figure}

We can initialize and solve the problem as before
\begin{minted}[mathescape]{python}
problem = Problem(geometry = geometry, n_fields = 2, self_adjoint = True,
                  L = L, 
                  Gamma = Gamma, 
                  F = F,
                  biases = [[1.,0.], [0.,0.]]) 
problem.solve()
\end{minted}
and make a plot of the voltage and temperature fields. These are reported in Figs. \ref{fig2}-\ref{fig3}.
\subsection{Post-processing}
\begin{figure}
\includegraphics[scale=0.75]{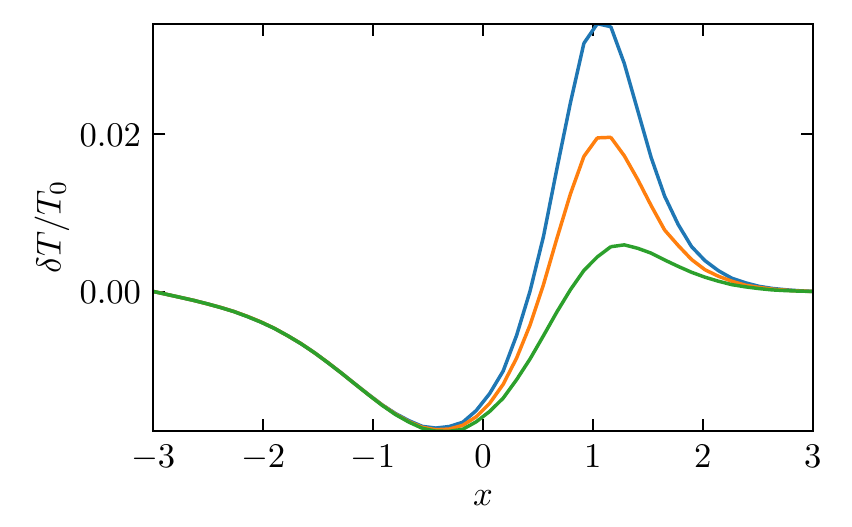}
\includegraphics[scale=0.75]{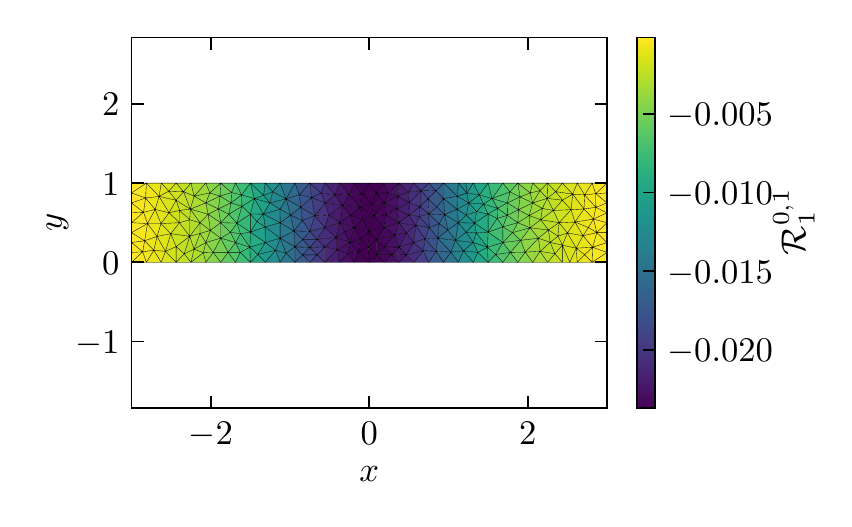}
\caption{\label{fig3} Left. Temperature field obtained for the thermoelctric problem represented as sections at $y=0$ (blue), $y=0.5$ (orange), and $y=1$ (green). Right. Generalized responsivity connecting a heat source to the charge current at contact 1.}
\end{figure}

In this case we also make a plot of the generalized responsivity ${\cal R}^{0,1}_1(\bm r)$ connecting a heat source to the charge current at contact 1.
\begin{minted}[mathescape]{python}
plt.axis('equal')
problem.plot_responsivity(flux = 0, contact = 1, source = 1)
cb = plt.colorbar()
cb.set_label('${\cal R}^{0,1}_1$')
plt.xlabel('$x$')
plt.ylabel('$y$')
plt.tight_layout()
\end{minted}
The result is shown in figure \ref{fig3} (right panel).

The responsivities are also available as Numpy arrays for further processing. The property \mintinline{python}{problem.responsivities_vertices} is a $N_{\rm f}\times N_{\rm c} \times N_{\rm f}\times N_{\rm vert}$ array that stores the vertex values of the responsivities.
More precisely the vertex values of ${\cal R}^{\alpha,m}_\beta$ are contained in \mintinline{python}{problem.responsivities_vertices[alpha,m-1,beta,:]}. Again, the vertex ordering is the same as in \mintinline{python}{problem.geometry.coordinates}.
\section{Summary and future development}
\label{summary}
In summary, we developed an easy-to-use FEM code, based on FEniCS, that allows the solution of coupled diffusion equations with matrix-valued coefficients in a realistic device geometries, with contact-like boundary conditions, including the impact of contact resistances.
We plan to extend the code to include non-linear effects in the form of dependences of the matrices $L_{\alpha\beta,ij}(\bm r)$ and $\Gamma_{\alpha\beta}(\bm r)$ on the fields and their gradients. This would allow the study of an even richer class of physical phenomena.  
Other future development will include non-diagonal contact resistances and the possibility of introducing anomalous contributions to the coefficients that are localized on lower-dimensional manifolds, to describe, for example contact resistances between different parts of a sample.
New version of the code will be available on the project repository \cite{repo} and made accessible via the Python Package Index.
\section*{Acknowledgements}
The author thanks Niels Hesp and Michele Torre for important feedback and suggestions, Jørgen S. Dokken for his invaluable help with the FEniCS library.
The calculation of generalized responsivities is based on an idea by M.B. Lundeberg.
The author acknowledges funding from the Spanish Ministry of Science, Innovation and Universities (MCIU) and
State Research Agency (AEI) via the Juan de la Cierva fellowship n. FJC2018-037098-I.
\appendix
\section{Proof of the uniqueness of the solution of Eq. (\ref{system})}
\label{proof_uniqueness}
Let $\phi_\alpha^{(1)}(\bm r)$ and $\phi_\alpha^{(2)}(\bm r)$ be two solutions of (\ref{system}) and $\psi_\alpha(\bm r) = \phi_\alpha^{(1)}(\bm r) - \phi_\alpha^{(2)}(\bm r)$ their difference. 
The function $\psi_\alpha(\bm r)$ respects (\ref{system}) with $F_\alpha(\bm r) \equiv 0$,
\begin{equation}\label{system_difference}
-\sum_{\beta ij}\partial_i[L_{\alpha\beta,ij}(\bm r)\partial_j\psi_\beta(\bm r)]
+\sum_{\beta}\Gamma_{\alpha\beta}(\bm r)\psi_\beta(\bm r)
=0.
\end{equation}
It also fulfils homogeneous Robin boundary conditions at the contacts 
\begin{equation}
\psi_\alpha(\bm r) + \bar{R}_{\alpha, n}\sum_{ij \beta} L_{\alpha\beta, ij}(\bm r)\partial_j\psi_\beta(\bm r) \hat{n}_i(\bm r)  = V_{\alpha, n} ,\quad \mbox{if } \bm r \in C_n,
\end{equation}
and homogeneous Neumann boundary conditions outside them.
Multiplying (\ref{system_difference}) by $\psi_\alpha(\bm r)$, summing over $\alpha$ and integrating over the volume we obtain
\begin{equation}
\int_\Omega d \bm r\sum_{\alpha\beta}\left[-\sum_{ij}\psi_\alpha(\bm r)\partial_i[L_{\alpha\beta,ij}(\bm r)\partial_j\psi_\beta(\bm r)]
+\psi_\alpha(\bm r)\Gamma_{\alpha\beta}(\bm r)\psi_\beta(\bm r)\right]
=0.
\end{equation}
Integrating the first term by parts and making use of the boundary condition yields
\begin{equation}\label{uniqueness}
\begin{split}
& \sum_n\sum_{\alpha\beta\gamma,ijk\ell}\bar{R}_{\alpha, n}\int_{C_n} d s \,\hat{n}_i\partial_i[L_{\alpha\beta,ij}(\bm r)\partial_j\psi_\beta(\bm r)]
\hat{n}_k\partial_k[L_{\alpha\gamma,k\ell}(\bm r)\partial_\ell\psi_\gamma(\bm r)]\\
&+\int_\Omega d \bm r\sum_{\alpha\beta}\left[\sum_{ij}\partial_i\psi_\alpha(\bm r)L_{\alpha\beta,ij}(\bm r)\partial_j\psi_\beta(\bm r)
+\psi_\alpha(\bm r)\Gamma_{\alpha\beta}(\bm r)\psi_\beta(\bm r)\right]
=0.
\end{split}
\end{equation}
Due to the fact that $L$ is positive defined and $\bar{R}$ and $\Gamma$ are semi-positive defined, the left-hand-side of (\ref{uniqueness}) is zero if and only if $\psi_\alpha(\bm r)$ vanishes identically. This proves that $\phi_\alpha^{(1)}(\bm r)$ and $\phi_\alpha^{(2)}(\bm r)$ cannot be different.
\section{Symmetry and signature of the response matrix}
\label{symmetry}
The response matrix can be rewritten as
\begin{equation}\label{symmetry_G}
\begin{split}
G_{\alpha,m;\beta,n} & = \int_{C_m}ds \sum_i \hat{n}_i(\bm r) J_{\alpha,i}^{(\beta,n)}(\bm r) = 
\sum_{p, \gamma} \int_{C_p}ds \sum_{i} \hat{n}_i(\bm r) J_{\gamma,i}^{(\beta,n)}(\bm r)\delta_{mp}\delta_{\alpha \gamma}\\
& =\sum_{p, \gamma} \int_{C_p}ds \sum_{i} \hat{n}_i(\bm r) J_{\gamma,i}^{(\beta,n)}(\bm r)[\tilde{\Phi}^{(\alpha,m)}_\gamma(\bm r) -\bar{R}_{\gamma, p}\sum_j\tilde{J}_{\gamma, j}^{(\alpha, m)}(\bm r)\hat{n}_j(\bm r)]\\
& =  \sum_\gamma\int_{\partial \Omega}ds \sum_{i} \hat{n}_i(\bm r) J_{\gamma,i}^{(\beta,n)}(\bm r)\tilde{\Phi}^{(\alpha,m)}_\gamma(\bm r)\\
&-\sum_{p,\gamma} \bar{R}_{\gamma, p}\int_{C_p}ds \sum_{i} \hat{n}_i(\bm r) J_{\gamma,i}^{(\beta,n)}(\bm r)\sum_j\tilde{J}_{\gamma, j}^{(\alpha, m)}(\bm r)\hat{n}_j(\bm r).
\end{split}
\end{equation}
Repeating the same steps on the response matrix of adjoint problem we obtain
\begin{equation}
\begin{split}
\tilde{G}_{\beta,n;\alpha,m} 
& =  \sum_\gamma\int_{\partial \Omega}ds \sum_{i} \hat{n}_i(\bm r) \tilde{J}_{\gamma,i}^{(\alpha,m)}(\bm r)\Phi^{(\beta,n)}_\gamma(\bm r)\\
&-\sum_{p,\gamma} \bar{R}_{\gamma, p}\int_{C_p}ds \sum_{i} \hat{n}_i(\bm r) J_{\gamma,i}^{(\beta,n)}(\bm r)\sum_j\tilde{J}_{\gamma, j}^{(\alpha, m)}(\bm r)\hat{n}_j(\bm r).
\end{split}
\end{equation}
The first term in (\ref{symmetry_G}) is also equal to
\begin{equation}
\begin{split}
& \sum_\gamma\int_{\partial \Omega}ds \sum_{i} \hat{n}_i(\bm r) J_{\gamma,i}^{(\beta,n)}(\bm r)\tilde{\Phi}^{(\alpha,m)}_\gamma(\bm r)=
\sum_\gamma\int_{\Omega}d {\bm r} \sum_{i} \partial_i[ J_{\gamma,i}^{(\beta,n)}(\bm r)\tilde{\Phi}^{(\alpha,m)}_\gamma(\bm r)]=\\
&= \sum_\gamma\int_{\Omega}d {\bm r} \sum_{i}[ \partial_i J_{\gamma,i}^{(\beta,n)}(\bm r)\tilde{\Phi}^{(\alpha,m)}_\gamma(\bm r)+J_{\gamma,i}^{(\beta,n)}(\bm r) \partial_i\tilde{\Phi}^{(\alpha,m)}_\gamma(\bm r)]\\
& =\sum_{\gamma\delta}\int_{\Omega}d {\bm r} [-\Gamma_{\gamma\delta}(\bm r)\Phi_{\delta}^{(\beta,n)}(\bm r)\tilde{\Phi}^{(\alpha,m)}_\gamma(\bm r)-\sum_{ij}L_{\gamma\delta,ij}(\bm r)\partial_j\Phi_\delta^{(\beta,n)}(\bm r) \partial_i\tilde{\Phi}^{(\alpha,m)}_\gamma(\bm r)]\\
& =\sum_{\gamma\delta}\int_{\Omega}d {\bm r}\Phi_\delta^{(\beta,n)}(\bm r) \{-\Gamma_{\delta\gamma}^T(\bm r)\tilde{\Phi}^{(\alpha,m)}_\gamma(\bm r)
+\sum_{ij}\partial_j[L_{\delta\gamma,ji}^T(\bm r) \partial_i\tilde{\Phi}^{(\alpha,m)}_\gamma(\bm r)]\}\\
&-\int_{\Omega}d {\bm r}\sum_{\gamma\delta,ij}\{\partial_j[\Phi_\delta^{(\beta,n)}(\bm r)L_{\delta\gamma,ji}^T(\bm r) \partial_i\tilde{\Phi}^{(\alpha,m)}_\gamma(\bm r)]\}\\
&=\sum_{\delta}\int_{\partial\Omega}ds\sum_j\hat{n}_j(\bm r)\Phi_\delta^{(\beta,n)}(\bm r)\tilde{J}_{\delta,j}^{(\alpha,m)}(\bm r),
\end{split}
\end{equation}
which proves
\begin{equation}
G_{\alpha,m;\beta,n} = \tilde{G}_{\beta,n;\alpha,m}.
\end{equation}
By following very similar step with $\Phi^{(\alpha,m)}_\gamma(\bm r)$ replacing $\tilde{\Phi}^{(\alpha,m)}_\gamma(\bm r)$ yields
\begin{equation}
\begin{split}
& G_{\alpha,m;\beta,n}   =  \sum_\gamma\int_{\partial \Omega}ds \sum_{i} \hat{n}_i(\bm r) J_{\gamma,i}^{(\beta,n)}(\bm r)\Phi^{(\alpha,m)}_\gamma(\bm r)\\
&-\sum_{p,\gamma} \bar{R}_{\gamma, p}\int_{C_p}ds \sum_{i} \hat{n}_i(\bm r) J_{\gamma,i}^{(\beta,n)}(\bm r)\sum_jJ_{\gamma, j}^{(\alpha, m)}(\bm r)\hat{n}_j(\bm r)\\
&  =-  \sum_{\gamma\delta}\int_{\Omega}d {\bm r} [\Phi^{(\alpha,m)}_\gamma(\bm r)\Gamma_{\gamma\delta}(\bm r)\Phi_{\delta}^{(\beta,n)}(\bm r)+\sum_{ij} \partial_i\Phi^{(\alpha,m)}_\gamma(\bm r)L_{\gamma\delta,ij}(\bm r)\partial_j\Phi_\delta^{(\beta,n)}(\bm r)]\\
&-\sum_{p,\gamma} \bar{R}_{\gamma, p}\int_{C_p}ds \sum_{i} \hat{n}_i(\bm r) J_{\gamma,i}^{(\beta,n)}(\bm r)\sum_j\hat{n}_j(\bm r)J_{\gamma, j}^{(\alpha, m)}(\bm r).
\end{split}
\end{equation}
Defining $\bar{\Phi}_\beta(\bm r)=\sum_{\alpha ,m}V_{\alpha, m}\Phi^{(\alpha,m)}_\beta(\bm r)$ and $\bar{J}_{\beta, i}(\bm r)=\sum_{\alpha ,m}V_{\alpha, m}J^{(\alpha,m)}_{\beta,i}(\bm r)$ as the corresponding current we can write
\begin{equation}
\begin{split}
& \sum_{\alpha\beta,mn}V_{\alpha, m}G_{\alpha,m;\beta,n} V_{\beta, n}=\\
&  =  -\sum_{\gamma\delta}\int_{\Omega}d {\bm r} [\bar{\Phi}_\gamma(\bm r)\Gamma_{\gamma\delta}(\bm r)\bar{\Phi}_{\delta}(\bm r)+\sum_{ij} \partial_i\bar{\Phi}_\gamma(\bm r)L_{\gamma\delta,ij}(\bm r)\partial_j\bar{\Phi}_\delta(\bm r)]+\\
&-\sum_{p,\gamma} \bar{R}_{\gamma, p}\int_{C_p}ds \sum_{i} \hat{n}_i(\bm r) \bar{J}_{\gamma,i}(\bm r)\sum_j\hat{n}_j(\bm r)\bar{J}_{\gamma, j}(\bm r)\leq 0.
\end{split}
\end{equation}
Where in the last step we used that $L$, $\Gamma$, and $\bar{R}$ are semi-positive defined.  
Since $L$ is also positive defined the equality sign holds only for vanishing $V_{\alpha, m}$.
\section{Proof of the reciprocity theorem Eq. (\ref{reciprocity})}
\label{app_reciprocity}
Starting from the definition (\ref{S_def}) we can make use of the Robin boundary conditions to obtain
\begin{equation}
\begin{split}
S_{\alpha m}[F] & = \int_{C_m}ds \sum_i \hat{n}_i(\bm r) J_{\alpha,i}^{F}(\bm r) = 
\sum_n \int_{C_n}ds \sum_{\beta, i} \hat{n}_i(\bm r) J_{\beta,i}^{F}(\bm r)\delta_{mn}\delta_{\alpha \beta}\\
& =\sum_n \int_{C_n}ds \sum_{\beta, i} \hat{n}_i(\bm r) J_{\beta,i}^{F}(\bm r)[\tilde{\Phi}^{(\alpha,m)}_\beta(\bm r) -\bar{R}_{\beta, n}\sum_j\tilde{J}_{\beta, j}^{(\alpha, m)}(\bm r)\hat{n}_j(\bm r)]\\
& =  \int_{\partial \Omega}ds \sum_{\beta, i} \hat{n}_i(\bm r) J_{\beta,i}^{F}(\bm r)\tilde{\Phi}^{(\alpha,m)}_\beta(\bm r)\\
&-\sum_{n,\beta} \bar{R}_{\beta, n}\int_{C_n}ds \sum_{i} \hat{n}_i(\bm r) J_{\beta,i}^{F}(\bm r)\sum_j\tilde{J}_{\beta, j}^{(\alpha, m)}(\bm r)\hat{n}_j(\bm r).
\end{split}
\end{equation}
The first term can be further elaborated making use of the adjoint PDE system to get
\begin{equation}
\begin{split}
&\sum_{\beta, i} \int_{\partial \Omega}ds  \hat{n}_i(\bm r) J_{\beta,i}^{F}(\bm r)\tilde{\Phi}^{(\alpha,m)}_\beta(\bm r)=
\sum_{\beta, i}\int_\Omega d{\bm r} \partial_i[J_{\beta,i}^{F}(\bm r)\tilde{\Phi}^{(\alpha,m)}_\beta(\bm r)]=\\
&=\sum_{\beta, i}\int_\Omega d{\bm r} [\partial_iJ_{\beta,i}^{F}(\bm r)\tilde{\Phi}^{(\alpha,m)}_\beta(\bm r) + J_{\beta,i}^{F}(\bm r)\partial_i\tilde{\Phi}^{(\alpha,m)}_\beta(\bm r)]=\\
&=\sum_{\beta}\int_\Omega d{\bm r} \{[F_\beta(\bm r)-\sum_\gamma\Gamma_{\beta\gamma}(\bm r)\Phi_{\gamma}^{F}(\bm r)]\tilde{\Phi}^{(\alpha,m)}_\beta(\bm r)+\\
& - \sum_{\gamma, ij} L_{\beta\gamma,ij}(\bm r)\partial_j \Phi_{\gamma}^{F}(\bm r)\partial_i\tilde{\Phi}^{(\alpha,m)}_\beta(\bm r)\}\\
&=\sum_{\beta}\int_\Omega d{\bm r} \{F_\beta(\bm r)\tilde{\Phi}^{(\alpha,m)}_\beta(\bm r)-\sum_\gamma\Phi_{\gamma}^{F}(\bm r)\Gamma_{\gamma\beta}^T(\bm r)\tilde{\Phi}^{(\alpha,m)}_\beta(\bm r)\\
& + \sum_{\gamma, ij} \Phi_{\gamma}^{F}(\bm r)\partial_j[L_{\gamma\beta,ji}^T(\bm r)\partial_i\tilde{\Phi}^{(\alpha,m)}_\beta(\bm r)]\}\\
&-\sum_{\beta}\int_\Omega d{\bm r}\sum_{\gamma, ij} \partial_j[ \Phi_{\gamma}^{F}(\bm r)L_{\gamma\beta,ji}^T(\bm r)\partial_i\tilde{\Phi}^{(\alpha,m)}_\beta(\bm r)]\\
&=\sum_{\beta}\int_\Omega d{\bm r} F_\beta(\bm r)\tilde{\Phi}^{(\alpha,m)}_\beta(\bm r)+\sum_{\gamma, j} \int_{\partial \Omega} ds\, \Phi_{\gamma}^{F}(\bm r)\hat{n}_j(\bm r)\tilde{J}^{(\alpha,m)}_{\gamma,j}(\bm r)\\
&=\sum_{\beta}\int_\Omega d{\bm r} F_\beta(\bm r)\tilde{\Phi}^{(\alpha,m)}_\beta(\bm r)+\sum_{n, \beta}\bar{R}_{\beta,n}\int_{C_n} ds\sum_{i} \hat{n}_i(\bm r) J_{\beta,i}^{F}(\bm r)\sum_{j} \hat{n}_j(\bm r)\tilde{J}^{(\alpha,m)}_{\beta,j}(\bm r).
\end{split}
\end{equation}
Substituting back into the previous equation yields
\begin{equation}
S_{\alpha m}[F] = \sum_{\beta}\int_\Omega d{\bm r} \tilde{\Phi}^{(\alpha,m)}_\beta(\bm r)F_\beta(\bm r).
\end{equation}
That comparing with (\ref{S_resp}) proves (\ref{reciprocity}).

An alternative way of rewriting $S$ is
\begin{equation}
\begin{split}
& S_{\alpha m}[F]
 =  \int_{\partial \Omega}ds \sum_{\beta, i} \hat{n}_i(\bm r) J_{\beta,i}^{F}(\bm r)\Phi^{(\alpha,m)}_\beta(\bm r)+\\
 &-\sum_{n,\beta} \bar{R}_{\beta, n}\int_{C_n}ds \sum_{i} \hat{n}_i(\bm r) J_{\beta,i}^{F}(\bm r)\sum_j J_{\beta, j}^{(\alpha, m)}(\bm r)\hat{n}_j(\bm r)\\
&=\sum_{\beta}\int_\Omega d{\bm r} \{\Phi^{(\alpha,m)}_\beta(\bm r)F_\beta(\bm r)-\sum_\gamma\Phi^{(\alpha,m)}_\beta(\bm r)\Gamma_{\beta\gamma}(\bm r)\Phi_{\gamma}^{F}(\bm r)+\\
& - \sum_{\gamma, ij} \partial_i\Phi^{(\alpha,m)}_\beta(\bm r) L_{\beta\gamma,ij}(\bm r)\partial_j \Phi_{\gamma}^{F}(\bm r)\}\\
&-\sum_{n,\beta} \bar{R}_{\beta, n}\int_{C_n}ds \sum_{i} \hat{n}_i(\bm r) J_{\beta,i}^{F}(\bm r)\sum_j J_{\beta, j}^{(\alpha, m)}(\bm r)\hat{n}_j(\bm r).
\end{split}
\end{equation}

\end{document}